\documentclass{Interspeech2024}

\usepackage{amsmath}
\usepackage{amssymb}
\usepackage{siunitx}
\usepackage{matvecsymb}
\usepackage{localmath}
\usepackage{IEEEtrantools}

\graphicspath{{./}{./figures/}}

\DeclareSIUnit\noop{\relax}

\renewenvironment{thebibliography}[1]{
  \begin{oldthebibliography}{#1}
  \setlength{\itemsep}{0.07em}
  \setlength{\parskip}{0.em}
}
{
  \end{oldthebibliography}
}

\interspeechcameraready

\title{Universal Score-based Speech Enhancement with High Content Preservation}

\name{Robin}{Scheibler}
\name{Yusuke}{Fujita}
\name{Yuma}{Shirahata}
\name{Tatsuya}{Komatsu}

\address{
  LY Corporation, Japan
\email{robin.scheibler@ieee.org}
}


\newcommand{\red}[1]{\textcolor{red}{#1}}

\begin{document}
\bstctlcite{IEEEexample:BSTcontrol}

\maketitle

\begin{abstract}
    We propose UNIVERSE++, a universal speech enhancement method based on score-based diffusion and adversarial training.
    Specifically, we improve the existing UNIVERSE model that decouples clean speech feature extraction and diffusion.
    Our contributions are three-fold.
    First, we make several modifications to the network architecture, improving training stability and final performance.
    Second, we introduce an adversarial loss to promote learning high quality speech features.
    Third, we propose a low-rank adaptation scheme with a phoneme fidelity loss to improve content preservation in the enhanced speech.
    In the experiments, we train a universal enhancement model on a large scale dataset of speech degraded by noise, reverberation, and various distortions.
    The results on multiple public benchmark datasets demonstrate that UNIVERSE++ compares favorably to both discriminative and generative baselines for a wide range of qualitative and intelligibility metrics.
\end{abstract}


\section{Introduction}

Speech enhancement (SE) is the task of restoring clean speech from a degraded signal~\cite{loizou_speech_nodate}.
Many SE methods operate on the magnitude spectrogram~\cite{boll_suppression_1979} with a mask estimated by statistical methods~\cite{ephraim_speech_1984} or, recently, deep neural networks (DNN)~\cite{tamura_noise_1988,lu_speech_2013,xu_regression_2015}.
See~\cite{lu_espnet-se_2022} for an overview of current methods.
DNN-based SE can be divided into time-domain~\cite{luo_conv-tasnet_2019} and time-frequency methods~\cite{yu_high_2023}.
Generally, the model is discriminatively trained to regress to the clean speech from the degraded signal.
However, this type of training is known to be prone to residual noise and artifacts~\cite{pirklbauer_evaluation_2023}.
An alternative approach is to use generative models, derived from text-to-speech technology, in the hope of generating high quality speech without residual noise.
Generative adversarial networks (GAN)~\cite{goodfellow_generative_2014} are suitable for SE.
The denoising network, playing the role of generator, is jointly trained with a discriminator that learns to distinguish between clean and enhanced speech~\cite{pascual_segan_2017,andreev_hifi_2023}.
Score-based diffusion~\cite{ho2020denoising,song_score-based_2022} is another generative paradigm that has recently gained traction for SE~\cite{lu_conditional_2022,richter_speech_2022,yen_cold_2023,serra_universal_2022}.
Diffusion-based SE models operate by progressively transforming normally distributed noise into clean speech conditioned on the input degraded speech.

This work focuses on universal speech enhancement (USE, also known as speech restoration), the extension of SE to all types of signal degradation, including reverberation, low-pass filtering, clipping, etc~\cite{serra_universal_2022,zhang_restoring_2021,byun_empirical_2023}.
It is believed that generative models are more suitable for USE as the model output is not uniquely determined by the input.
For example, there are many plausible reconstructions for a low-pass signal.
UNIVERSE~\cite{serra_universal_2022} is a diffusion-based method for USE that attracted attention due to a high profile demonstration featuring very high quality enhanced speech.
However, during preliminary experiments, we found it difficult to train, and prone to hallucinations such as mumbled speech.
We address these issues in the UNIVERSE++ model with a number of improvements over the original.
First, we introduce several architectural upgrades such as normalization and anti-aliasing filters in the down/up-sampling layers of the network.
Second, we combine the score matching training with the adversarial loss of HiFi-GAN~\cite{jungil2020hifigan} that promotes the extraction of high-quality features of the clean speech to condition the diffusion process.
Finally, we propose a light-weight low-rank adaptation (LoRA)~\cite{hu_lora_2022} procedure to improve the retention of linguistic content in the enhanced speech.
Extensive experiments on multiple benchmarks demonstrate the versatility and effectiveness of the proposed method.
In particular, we show that it produces highly natural speech without compromising on the intelligibility of the content.
Our implementation of UNIVERSE++ is available as open-source\footnote{\protect{\href{https://github.com/line/open-universe}{\texttt{https://github.com/line/open-universe}}}}.



\section{Background}

The task of \textit{universal} speech enhancement (USE) consists in recovering a clean speech signal $\vx$ that is only available through a degraded signal $\vy$.
Unlike conventional SE, the degradation is not limited to additive noise, but may include reverberation, clipping, filtering, coding artifacts, etc.
Throughout the paper, bold upper and lower case letters are for matrices and vectors, respectively.
The Euclidean norm of $\vv\in\R^n$ is $\lVert \vv \rVert = (\vv^\top \vv)^{\half}$.
UNIVERSE has been proposed to tackle this task~\cite{serra_universal_2022}.
It is composed of two networks, as shown in \ffref{network}.
A conditioning network $C(\,.\,)$ extracts clean speech features from the degraded speech.
Then, these features are used to condition a diffusion process synthesizing the clean speech directly.
The second network is the score denoising model $S(\,.\,)$ used as part of the diffusion process.
This division of labor in the network makes it easy to add losses to the conditioning network to control the extracted features.

UNIVERSE follows the stochastic differential equation (SDE) formulation of the diffusion models~\cite{song_score-based_2022}.
This framework defines a continuous time process transforming the clean speech $\vx_0$ into normally distributed noise until the original signal is imperceptible.
According to theory of SDE, under mild conditions, the reverse process exists and depends only on the so-called \textit{score function}, the derivative of the logarithm of the marginal distribution at time $t$ of the process.
Although unknown in practice, a neural network can be trained to approximate the score function by minimizing the following objective,
\begin{align}
  \calL_{\text{score}} = \E_{t,\vz,\vx}\left[ \lVert \sigma_t S(\vx + \sigma_t \vz, \sigma_t, \vc) + \vz \rVert^2 \right],
  \label{eq:loss_score}
\end{align}
where $\sigma_t^2$ is the noise variance at time $t\in[0,1]$, and $\vz\sim\calN(\vzero, \mI)$.
Furthermore, $\vc = C(\vy)$ is the feature vector produced by the conditioning network.
The noise variance follows an exponential schedule $\sigma_t^2 = \sigma_{\min}^2 (\sigma^2_{\max}/\sigma^2_{\min})^t$, where $\sigma_{\min}$ and $\sigma_{\max}$ are hyperparameters.
At inference time, the estimated clean speech is produced by solving the reverse diffusion process with the noise-consistent Langevin dynamics~\cite{jolicoeur-martineau2021adversarial},
\begin{align}
  \vx_{t-\Delta} = \vx_t + \eta \sigma_t^2 S(\vx_t, \vc, \sigma_t) + \beta \sigma_{t-\Delta} \vz.
\end{align}
For $N$ steps, $\Delta = \nicefrac{1}{N}$. The parameters $\eta = 1 - \gamma^\epsilon$ and $\beta = \sqrt{1 - \gamma^{2(\epsilon - 1)}}$ with $\gamma = \nicefrac{\sigma_{\max}}{\sigma_{\min}}$, and $\epsilon$ is a hyperparameter.

\begin{figure}
  \centering
  \includegraphics[width=\linewidth]{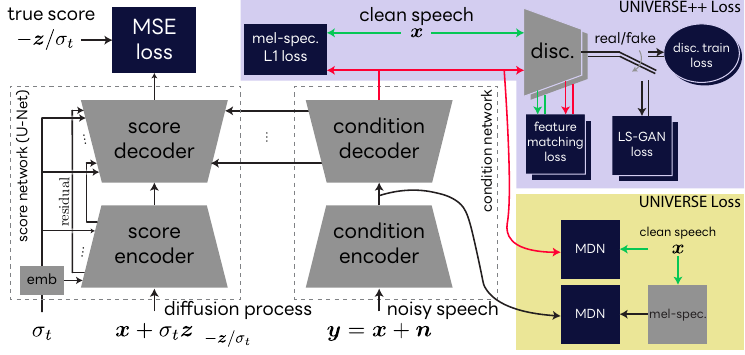}
  \caption{Overview of the UNIVERSE network architecture with proposed (purple box) and original (yellow box) loss functions.}
  \flabel{network}
\end{figure}

\textbf{Model Architecture} The two networks share a similar encoder/decoder structure, each with four stages, that gradually reduce/increase the sampling rate and increase/reduce the number of channels.
Each stage is composed of three convolutional blocks, one of which changes the sampling rate.
The convolutional blocks are interleaved with PReLU activation functions.
Gated recurrent units are used to model the temporal dependencies in the bottleneck of the network with two and one layers in the conditioning and score networks, respectively.
The intermediate features produced at each stage of the decoder in the conditioning network are injected in the corresponding stage of the score network via linear projection layers.
In the score network, residual connections are used between corresponding stages of the encoder and decoder.
In the conditioning networks, residual connections connect each stage of the encoder to the bottleneck features via a convolutional adaptation layer.
The noise variance of the diffusion process is injected at all stages of the score network via a random Fourier embedding and FiLM layers.
See \ffref{network} for an overview, and~\cite{serra_universal_2022} for the full details.

\textbf{Loss Functions and Training}
In addition to the score matching loss~\eqref{eq:loss_score}, UNIVERSE uses a number of mixture density network (MDN) losses~\cite{bishop_mixture_1994} to promote the learning of good features.
One MDN matches the bottleneck features of $C$ to the clean speech mel-spectrogram.
Another, connected to the last layer of the decoder, matches the output features to the clean speech waveform.
In the original UNIVERSE, a number of other losses with respect to speech features such as pitch and voice activity and other speech features are also used.
However, we did not consider them as they were not shown to improve the metrics in~\cite{serra_universal_2022}.
The model was trained on a large private dataset of speech and noise coupled to dynamic distortions~\cite{serra_universal_2022}.

\section{UNIVERSE++}
\seclabel{universe++}

Our modifications to the UNIVERSE methodology are three-fold.
First, we introduce improvements in the network to follow best practices for diffusion and synthesis.
Second, we introduce adversarial training to improve the quality of the extracted features.
Finally, we propose a fine-tuning procedure directly targeting the linguistic content.

\subsection{Proposed Network Improvements}

\textbf{Normalization} Due to the exponential noise schedule in the diffusion process, the score network has to deal with inputs and outputs with widely different scales spanning up to four orders of magnitude.
This issue is tackled by Karras et al.~\cite{karras_elucidating_2022} who propose to re-parameterize the score network as
\begin{align}
  S(\vx, \vc, \sigma_t) &= c_t^{\text{skip}}\vx + c_t^{\text{out}} S^\prime\left( c_t^{\text{in}} \vx, \vc, \sigma_t \right),
\end{align}
where the weights depending on the variance at time $t$,
\begin{align}
  c_t^{\text{skip}}=\frac{\sigma_{\text{data}}^2 }{ \sigma_{\text{data}}^2 + \sigma_t^2},
  \ 
  c_t^{\text{out}}=\sigma_t \sqrt{c_t^{\text{skip}}},
  \ 
  c_t^{\text{in}}=1/\sqrt{\sigma_{\text{data}}^2 + \sigma_t^2},
  \nonumber
\end{align}
are such that the network inputs and training targets have unit variance.
The $\sigma_{\text{data}}$ is the variance of the clean speech data.

\noindent \textbf{Anti-aliasing}
In the score network UNet, down- and up-sampling stages introduce aliasing artifacts which have been show to be detrimental to translation invariance in image generation networks~\cite{karras_alias-free_2021}.
Following the practice of image generation~\cite{karras_alias-free_2021}, we introduce anti-aliasing filters before any change of sampling rate.
As shown in \ffref{antialiasing}, the high-frequency content is propagated only via the residual connection and its processing is cleanly restricted to the upper stages of the network.
We did not use antialiasing in the conditioning network, that lacks residual connections, as we observed a performance degradation.

\begin{figure}
  \centering
  \includegraphics[width=\linewidth]{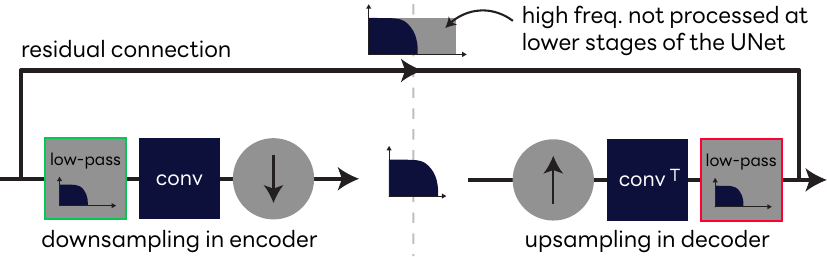}
  \caption{Proposed anti-aliasing filters at down/up-sampling stages of the score network UNet.}
  \flabel{antialiasing}
\end{figure}

\noindent \textbf{Miscellaneous}
We use the weight normalization parameterization~\cite{salimans_weight_2016} for the convolutional and linear layers.
The original embeddings for $\sigma_t$ use random Fourier features with a three layer multi-layer perceptron (MLP).
Since the embeddings are followed by yet another projection layer, the MLP seemed redundant and we use non-random Fourier embedding instead
\begin{align}
  \ve(\sigma_t) = \left[ \cdots\, \cos(2\pi f(\sigma_t)m)\, \cdots\, \sin(2\pi f(\sigma_t) m)\, \cdots \right]^\top.
  \nonumber
\end{align}
where $f(\sigma_t) = \alpha \log(\sigma_t) + \beta$, $\alpha,\beta$ are trainable.

\subsection{HiFi-GAN Adversarial Loss}
Despite its probabilistic formulation, the original MDN loss is still applied sample-wise to the target speech in a discriminative fashion.
Instead, we replace it by the adversarial loss of HiFi-GAN~\cite{jungil2020hifigan} (purple box in \ffref{network}).
It uses two discriminators, multi-period and multi-resolution, trained with the least-squares GAN loss~\cite{mao_least_2017} .
A mel-spectrogram loss between the output of the adapter and the clean speech improves the training efficiency and the fidelity of the features produced.
Finally, feature matching losses between the discriminators' feature maps of a clean sample and the output of the conditioning network allow to capture high-level features of the clean speech content.

\subsection{Linguistic Content Enhancing Low-rank Adaptation}

As reported in~\cite{pirklbauer_evaluation_2023,scheibler_diffusion-based_2023}, we noticed that while the generatively enhanced speech sounds very natural, hallucinations tend to appear in low signal-to-noise ratio (SNR) segments.
We address this issue with a fine-tuning procedure to improve the retention of linguistic content.
We select LoRA for the fine-tuning~\cite{hu_lora_2022}.
Due to its low-memory footprint, we can perform the fine-tuning on the inference stage of the diffusion process.
We back-propagate through the last two steps, similarly to~\cite{lay_reducing_2023}, and the conditioning network.
We use a phoneme prediction model~\cite{xu_simple_2022}, which is language agnostic, and avoids requiring transcripts of the training data, to predict the phonemes of the clean and enhanced speech.
A connectionist temporal classification (CTC) loss~\cite{graves_connectionist_2006} is used between the predicted phonemes of the clean speech and those of the enhanced speech.
We note that phoneme losses have been used with text labels for speech synthesis~\cite{zhang_visinger_2022}.
The speech integrity is preserved by the frozen Hifi-GAN losses, and a multi-resolution spectrogram loss.

\begin{figure}
  \centering
  \includegraphics{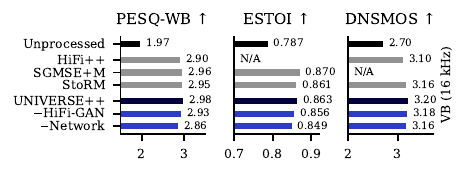}
  \caption{Results on the Voicebank-DEMAND dataset at \SI{16}{\kilo\hertz}. The values for HiFi++ and SGMSE+M are those reported in~\cite{andreev_hifi_2023} and~\cite{lemercier_storm_2023}, respectively.}
  \flabel{vb_results}
\end{figure}

\section{Experiments}

First, we evaluate UNIVERSE++ for additive noise reduction on the Voicebank+DEMAND (VB) dataset~\cite{valentini-botinhao_speech_2016}, enabling comparison to a large number of SE methods.
We also perform an ablation study to evaluate the gains due to architecture and loss function.
Our second experiment studies the performance of UNIVERSE++ for USE on several public benchmark datasets.

\subsection{Evaluation Metrics}

Generative methods in SE require the use of a variety of metrics~\cite{pirklbauer_evaluation_2023}.
The waveform reconstruction is measured by PESQ~\cite{rix_perceptual_2001} and log-spectral distance (LSD), and intelligibility by extended STOI~\cite{jensen_algorithm_2016}.
To evaluate the linguistic content preservation of the algorithms, we use the Levenshtein phoneme similarity (LPS), i.e., the phoneme accuracy of the enhanced versus clean speech according to a phoneme predictor~\cite{pirklbauer_evaluation_2023}, and the word error rate (WER) of the Whisper large-v3 speech recognition model~\cite{radford_robust_2023}.
Naturalness is measured with the non-intrusive neural mean opinion score DNSMOS~\cite{reddy_dnsmos_2022}.

\subsection{Noise Reduction on Voicebank+DEMAND}
\seclabel{vb}

\textbf{Dataset} The Voicebank+DEMAND dataset is originally split into train and test~\cite{valentini-botinhao_speech_2016}.
Following standard practice, we form a validation set with speakers p226 and p287 of the training set.
This results in \SI{8.75}{\hour} for training, \SI{38}{\minute} for validation, and \SI{34}{\minute} for testing.
The sampling frequency is \SI{16}{\kilo\hertz}.

\noindent \textbf{Training} We train the model with the AdamW optimizer~\cite{loshchilov_decoupled_2018}.
The learning rate starts at $10^{-6}$ and linearly increases to $10^{-4}$ over the first 10000 steps.
A cosine schedule reduces the learning rate to $10^{-6}$ from steps 200000 to 300000 when the training stops.
The batch size is 40.
An exponential moving average of the weights with forgetting factor \num{0.999} is used for inference.

\noindent \textbf{Baselines} HIFI++ is a generative method enhancing the mel-spectrogram and using a HiFi-GAN neural vocoder to reconstruct the clean speech from it~\cite{andreev_hifi_2023}.
SGMSE+M is a generative speech enhancement method using diffusion based on the Ornstein-Uhlenbeck SDE~\cite{lemercier_storm_2023}.
StoRM is a hybrid method that uses a jointly, but discriminatively, trained network to initialize the diffusion process of SGMSE+M~\cite{lemercier_storm_2023}.

\noindent \textbf{Results} \ffref{vb_results} shows that omitting from UNIVERSE++ the adversarial loss ($-$HiFi-GAN) and further removing the network improvements ($-$Network) reduce PESQ by 0.05 and 0.07 points, respectively.
UNIVERSE++ outperforms the baseline methods for PESQ and DNSMOS, while being less than 0.01 point STOI behind StoRM.

\begin{figure*}
  \centering
  \includegraphics{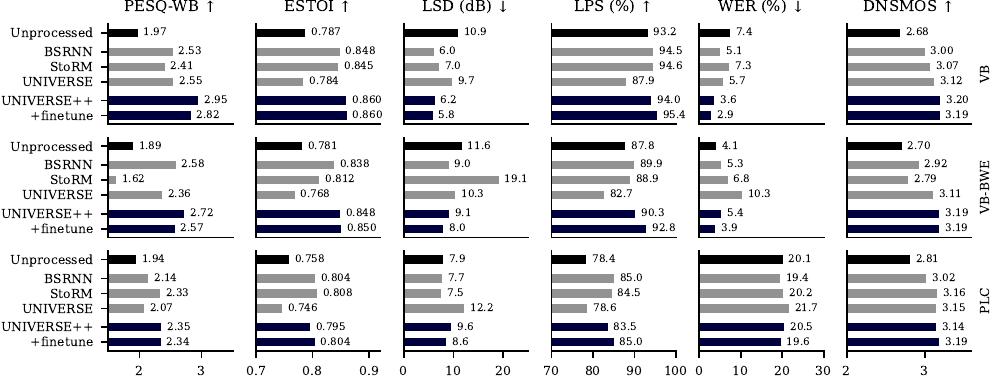}
  \caption{Results on the Voicebank+DEMAND (VB) dataset, VB low-pass filtered at \SI{4}{\kilo\hertz} (VB-BWE), and packet loss concealment challenge validation set (PLC). Arrows indicate if the metric is better when increasing ($\uparrow$) or decreasing ($\downarrow$).}
  \flabel{universal_results}
\end{figure*}

\subsection{Universal Speech Enhancement}

\subsubsection{Datasets}
\seclabel{universal_dataset}

\textbf{Training} We train our universal enhancement model at \SI{24}{\kilo\hertz} on a large scale dataset of speech degraded by noise, reverberation, and various distortions applied dynamically during training.
The clean speech dataset is composed of the following sources.
The training set of the URGENT challenge~\cite{urgent_2024}, which was itself created from the English read speech part of the deep noise suppression (DNS) challenge~\cite{dubey_icassp_2022}, the Common Voice English v11~\cite{ardila_common_2020}, LibriTTS~\cite{zen_libritts_2019}, and VCTK~\cite{veaux_voice_2013}.
The dataset was filtered to remove samples originally collected at less than \SI{24}{\kilo\hertz} or of low quality, as evaluated using the DNSMOS metric~\cite{reddy_dnsmos_2022}, resulting in \SI{300}{\hour} of high quality speech.
See~\cite{urgent_2024} for the full details.
In addition, we use an internal dataset of \SI{237}{\hour} of studio quality Japanese speech from 17 speakers.
This totals \SI{537}{\hour} of clean speech.
The noise dataset consists of \SI{177}{\hour} from Audio Set~\cite{gemmeke_audio_2017} (from the DNS challenge training set), \SI{78}{\hour} from the Wham! dataset~\cite{Wichern2019WHAM}, and \SI{346}{\hour} of an internal background music dataset.
The noise is mixed with an SNR uniformly distributed between \SIrange{-5}{30}{\decibel}, except for the background music for which it is between \SIrange{-10}{5}{\decibel}.
We dynamically apply one of the 132037 real room impulse responses of the Arni~\cite{prawda_karolina_2022_6985104} dataset, or of 100000 synthesized using Pyroomacoustics~\cite{Scheibler:2018di}.
We also apply band limitation, equalization distortion, clipping (clamp, tanh, sigmoid), random attenuation, packet loss, or codec distortion (MP3).
One to five types of distortions are applied at random following approximately the methodology of UNIVERSE~\cite{serra_universal_2022}.
The two internal datasets were necessary for our target downstream task.

\begin{figure}
  \centering
  \includegraphics{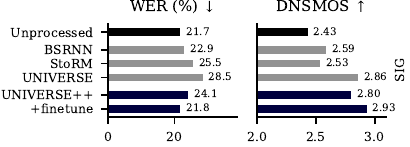}
  \caption{Results on the Signal Improvement Challenge non-blind test set, for which reference clean speech is not available.}
  \flabel{universal_results_sig}
\end{figure}

\medskip

We use the following test sets, all resampled at \SI{24}{\kilo\hertz}.

\noindent \textbf{Voicebank+DEMAND (VB)} This is the test split of the VB dataset described in \sref{vb} but sampled at \SI{24}{\kilo\hertz}.

\noindent \textbf{VB + Bandwidth Extension (VB-BWE)} Same as above, but we apply in addition a low-pass filter with cut-off at \SI{4}{\kilo\hertz} to test the bandwidth extension capability of the model.

\noindent \textbf{Packet Loss Concealment (PLC)} Validation set of the ICASSP 2024 Audio Deep Packet Loss Concealment Challenge~\cite{diener_icassp_nodate}.
The dataset contains 800 samples of speech with burst of packet loss of length up to \SI{3}{\second} realistically distributed.

\noindent \textbf{Signal Improvement Challenge (SIG)} ICASSP 2023 Speech Signal Improvement Challenge test set~\cite{cutler_icassp_2023}.
The dataset contains 500 samples of real-world recorded distorted speech in five different languages.
It includes various distortions such as noise (background, circuit, or coding), reverberation, clipping, etc.

\subsection{Model Details and Training}

The rate change factors in the four stages of the encoder/decoder are set to 2, 3, 5, 8 so that the rate at the bottleneck layer is \SI{10}{\hertz} as in~\cite{serra_universal_2022}.
The initial number of channels is 48.
The model has \SI{107.5}{\mega\noop} parameters and is trained for 1500000 steps with the AdamW optimizer~\cite{loshchilov_decoupled_2018}.
The HiFi-GAN losses add \SI{41.4}{\mega\noop} parameters during training.
The learning rate linearly goes from $10^{-6}$ to $10^{-4}$ over the first 50000 steps, then back to $10^{-6}$ with a cosine schedule over the last 500000.
The batch size is 40.
The training samples are truncated to \SI{2}{\second}.
During fine-tuning, the rank of the adaptation matrix is 16.
Matrices with either dimension smaller are not updated.
The learning rate is $10^{-5}$ and the samples are truncated to \SI{4}{\second}.
The batch size is 16.
The number of diffusion steps for inference is 8 and $\epsilon=1.3$.

\subsection{Baselines}

\textbf{Band-split recurrent neural network (BSRNN)} is a strong baseline for discriminative SE~\cite{yu_high_2023}.
We train a model with 128 channels and 6 layers.
The STFT uses window size and shift of \SI{20}{\milli\second} and \SI{5}{\milli\second}, respectively.
We split the spectrum into 10, 12, 8, and 4 bands in the range of \SIrange{0}{1}{\kilo\hertz}, \SIrange{1}{4}{\kilo\hertz}, \SIrange{4}{8}{\kilo\hertz}, and \SIrange{8}{12}{\kilo\hertz}, respectively.
The number of parameters is \SI{11.5}{\mega\noop}.
We trained for 750000 steps with batch size 48 and the loss of~\cite{yu_high_2023}.
The checkpoint with highest validation PESQ was selected.

\noindent \textbf{StoRM} is trained using the code of \cite{lemercier_storm_2023} modified to work with the dataset of \sref{universal_dataset}. 
The model has \SI{39.2}{\mega\noop} trainable parameters.
The model is trained for 1500000 steps with a batch size of 32.
Inference is done with 50 diffusion steps.

\noindent \textbf{UNIVERSE} is our re-implementation of~\cite{serra_universal_2022} without the proposed modifications of \sref{universe++} (\SI{107.7}{\mega\noop} parameters and another \SI{7.6}{\mega\noop} for the MDN losses during training).

\subsection{Results}

The results of evaluation on the tests sets are shown in \ffref{universal_results} and~\ref{fig:universal_results_sig}.
The purely discriminatively trained BSRNN has high content retention demonstrated by low LSD, low WER, and high LPS on all datasets.
However, it is markedly worse in terms of naturalness, demonstrated by the low DNSMOS (OVRL) values.
StoRM exhibits attributes of both generative and discriminative methods with high naturalness, as well as good overall content preservation.
However, its performance is not consistent over all datasets, e.g., it does not do well on the bandwidth extension task.
The UNIVERSE baseline has higher DNSMOS than BSRNN, but is overall worse than all other methods.
High LSD and WER, and low STOI and LSD suggest it may change content or drop segments, which is confirmed by informal listening.
The improved UNIVERSE++ is the most versatile in that it performs well with respect to all metrics and on all datasets.
It does not damage the linguistic content, as shown by low WER at the same level, or below, that of the unprocessed speech.
Fine-tuning further improves the WER and LPS, at the expense of some PESQ.
Anecdotally, UNIVERSE(++) is much faster than StoRM, as it only requires 8 diffusion steps compared to 50 for StoRM.

\section{Conclusion}

We proposed UNIVERSE++, a universal speech enhancement method using score-based diffusion and adversarial training.
Experiments demonstrate that the proposed model improves over the original UNIVERSE and also outperforms conventional methods on several test sets covering a wide range of speech distortions.
The adversarial loss significantly improves both the naturalness of the enhanced speech and the linguistic content preservation.
Together with the fine-tuning procedure, the method achieves the same performance as discriminative methods in terms of content preservation, but produces significantly more natural enhanced speech.
In future work, we will explore the use of phoneme loss during the main training stage.

\bibliographystyle{IEEEtran}
\bibliography{refs}

\end{document}